\documentclass{article}

\usepackage{arxiv}

\usepackage[utf8]{inputenc} 
\usepackage[T1]{fontenc}    
\usepackage{hyperref}       
\usepackage{url}            
\usepackage{booktabs}       
\usepackage{amsfonts}       
\usepackage{nicefrac}       
\usepackage{microtype}      
\usepackage{lipsum}		
\usepackage{graphicx}
\usepackage{natbib}
\usepackage{doi}
\usepackage{pifont}
\usepackage[table]{xcolor}
\definecolor{lightgray}{gray}{0.9}
\definecolor{red}{rgb}{0.95, 0.0, 0.24}
\definecolor{green}{rgb}{0.0, 0.5, 0.0}
\usepackage{booktabs}

\title{ReservoirComputing.jl: An Efficient and Modular Library for Reservoir Computing Models}

\author{ \href{https://orcid.org/0000-0000-0000-0000}{\includegraphics[scale=0.06]{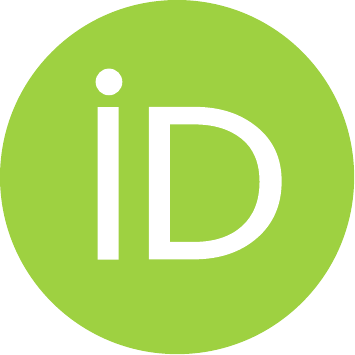}\hspace{1mm}Francesco Martinuzzi} \\
	Remote Sensing Centre for Earth System Research\\
	Center for Scalable Data Analytics and Artificial Intelligence\\
	Julia Computing Inc.\\
	\texttt{francesco.martinuzzi@uni-leipzig.de} \\
	\And
	\href{https://orcid.org/0000-0001-5850-0663}{\includegraphics[scale=0.06]{orcid.pdf}\hspace{1mm}Chris Rackauckas} \\
	Massachusetts Institute of Technology\\
	Julia Computing Inc.\\
	\texttt{crackauc@mit.edu} \\
	\And
	\hspace{1mm}Anas Abdelrehim \\
	Julia Computing Inc.\\
	\texttt{anasabdel.rehim@juliacomputing.com} \\
	\And
	\href{https://orcid.org/0000-0003-3031-613X}{\includegraphics[scale=0.06]{orcid.pdf}\hspace{1mm}Miguel D. Mahecha} \\
	Remote Sensing Centre for Earth System Research\\
	\texttt{miguel.mahecha@uni-leipzig.de} \\
	\And
	\href{https://orcid.org/0000-0002-3323-4490}{\includegraphics[scale=0.06]{orcid.pdf}\hspace{1mm}Karin Mora} \\
	Remote Sensing Centre for Earth System Research\\
	German Centre for Integrative Biodiversity Research (iDiv) \\
	\texttt{karin.mora@uni-leipzig.de} \\
}

\date{}

\renewcommand{\shorttitle}{ReservoirComputing.jl}

\hypersetup{
pdftitle={ReservoirComputing.jl: An Efficient and Modular Library for Reservoir Computing Models},
pdfsubject={cs.MS, cs.AI},
pdfauthor={Francesco Martinuzzi, Chris Rackauckas, Anas Abdelrehim, Miguel Mahecha and Karin Mora},
pdfkeywords={reservoir computing, echo state networks, julia, machine learning},
}

\begin{document}
\maketitle

\begin{abstract}
We introduce ReservoirComputing.jl, an open source Julia library for reservoir computing models. The software offers a great number of algorithms presented in the literature, and allows to expand on them with both internal and external tools in a simple way. The implementation is highly modular, fast and comes with a comprehensive documentation, which includes reproduced experiments from literature. The code and documentation are hosted on Github under an MIT license \url{https://github.com/SciML/ReservoirComputing.jl}.
\end{abstract}

\keywords{reservoir computing, echo state networks, julia, machine learning}

\section{Introduction}

Time series modeling is a very common technique throughout many areas of machine learning. 
However, many standard recurrent models are known to be susceptible to problems such as the vanishing gradient \citep{pascanu2013difficulty} or the extreme sensitivity of chaotic systems to their parameterization \citep{wiggins2003introduction}. To counter these issues reservoir computing (RC) techniques were introduced as recurrent models which can be trained without requiring gradient-based approaches \citep{lukovsevivcius2009reservoir}. 
Independently proposed as echo state networks (ESNs) \citep{jaeger2001echo} and liquid state machines (LSMs) \citep{maass2002real}, these architectures are based on the expansion of the input data using a fixed random internal layer, known as the reservoir, and the subsequent mapping of the reservoir to match an output. This output mapping can normally be written in a simple closed form, such as an $L_2$ minimization computed via a single QR-factorization. This allows the RC approach to achieve faster computational times with less parameter tuning compared to deep learning models which rely on local optimization. Numerous alterations have been made since their inception, such as deep architectures \citep{gallicchio2018design} and different training approaches \cite{chatzis2011echo, shi2007support}. Applications have confirmed that these improved architectures and training techniques can be impactful in many fields of study: 
from surrogates for stiff systems \citep{anantharaman2020accelerating} to reconstruction of chaotic attractors \citep{pathak2018model} and prediction of climate dynamics \citep{nadiga2021reservoir}. 
This expanding landscape necessitates the development of production-quality software to empower scientists with all of the latest techniques without having to implement from scratch every novel variation on their own. While a number of libraries for ESNs and RC are available \citep{trouvain2020reservoirpy, steiner2021pyrcn, pontes2020neuro}, they lack the hackability necessary to alter the provided architectures and achieve the top performances from the literature. To provide a truly modular library for RC models we present ReservoirComputing.jl, an efficient and flexible library written using the Julia language \citep{bezanson2017julia}. This software provides a user oriented package with intuitive high level application programming interfaces (APIs), while maintaining a great level of customization. This flexibility is obtained from both the design choices of ReservoirComputing.jl and by leveraging the multiple dispatch architecture of Julia. This gives users the freedom to mix any Julia code
with the reservoir computers in a seamless way to promote the latest performance and research.

\section{Theoretical Background}
\label{theo}
The setup of a RC model \citep{Konkoli2018} is based on leveraging the complex dynamics of a nonlinear system, the reservoir $\mathcal{R}$. The role of this layer is to expand the input data $\mathbf{u}(t)$ at time $t$ into a higher dimension. This operation can be aided by an input layer $\chi$, which maps the data onto the reservoir. The input data $\mathbf{u}(t)$ drives the reservoir $\mathcal{R}$ dynamics to a certain configuration. The resulting state $\mathbf{x}(t)$ represents the expansion of the input data and is used for training. Once the states have been obtained for all the training length $T$ the RC model can be trained. The states $\mathbf{x}(t)$ can be manipulated at this stage: through concatenation with the input data $\mathbf{u}(t)$ such as $\mathbf{z}(t)=[\textbf{x}(t), \textbf{u}(t)]$ \citep{Jaeger:2007}, using padding with some constant value $c$ usually set $c=1.0$ such as $\mathbf{z}(t)=[c, \textbf{u}(t)]$ \citep{lukovsevivcius2012practical}, or through nonlinear transformations \citep{chattopadhyay2020data}. The training is performed in one step in order to find the best fit between the states $\mathbf{x}(t)$ and the desired output $\mathbf{y}(t)$g. The most common way to perform the training is linear regression, creating the readout layer $\psi$. 
This layer is finally used in the prediction phase to obtain the desired output $\mathbf{v}(t) = \psi (\mathbf{x}(t))$.

\section{Library Overview}

\begin{figure}[t]
  \centering
  \includegraphics[width=1.0\textwidth]{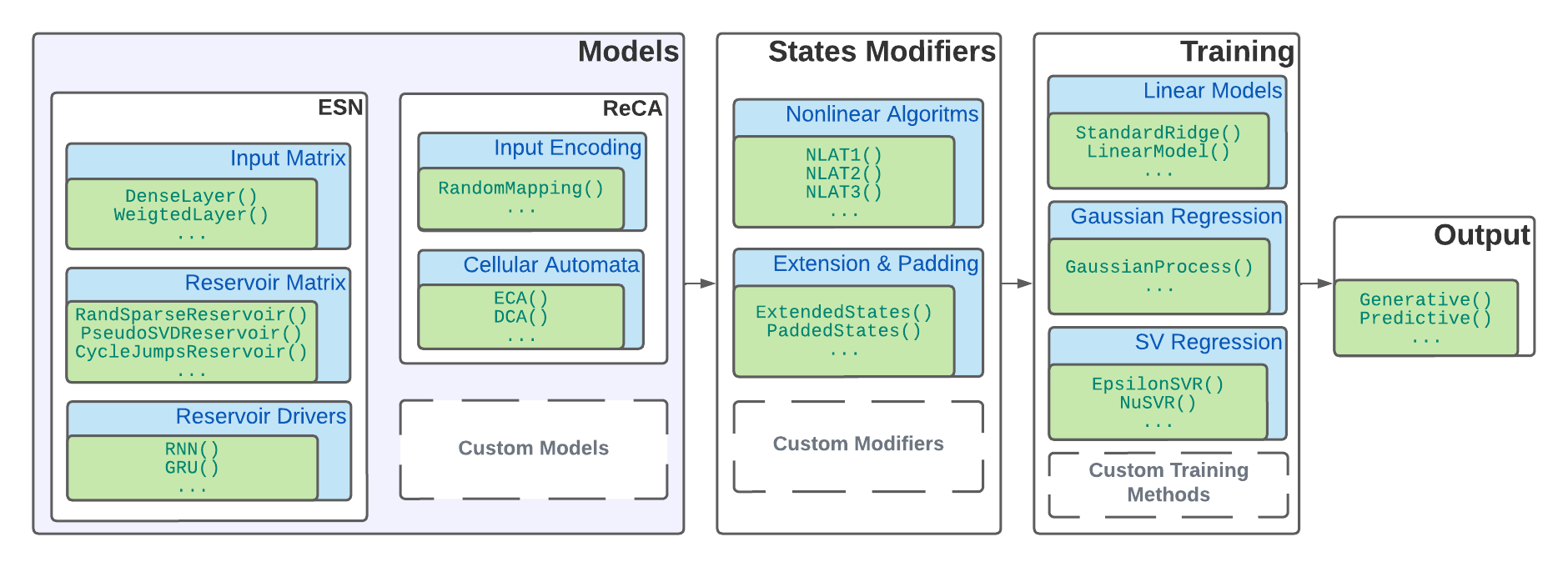}
  \caption{ReservoirComputing.jl architecture. \emph{SV} stands for support vector.}
  \label{fig: rc}
\end{figure}

ReservoirComputing.jl features the implementation of different RC models, training methods, and state variations. The architecture with the main components are detailed below and illustrated in Figure \ref{fig: rc}. Given the modular nature of the software 
more algorithms can be easily implemented on top of the existing ones. Any of the many libraries of the Julia community or custom codes can be used as part of the functions within the reservoir computer. Importantly, as Julia is a just-in-time (JIT) compiled language, these extensions do not sacrifice performance.

\paragraph{Building models.} ReservoirComputing.jl provides a simple interface for model building that closely follows the workflow presented in the corresponding literature. It includes a standard implementation of ESNs \citep{lukovsevivcius2012practical} as well as a hybrid variation \citep{pathak2018hybrid}, gated recurrent unit ESN \citep{wang2020gated} and double activation function ESN \citep{lun2015novel}. Multiple input layers $\chi$ and reservoirs $\mathcal{R}$ are also provided, ranging from weighted input layers \citep{lu2017reservoir} to minimally complex input layers and reservoirs \citep{rodan2010minimum, rodan2012simple}, including a reservoir obtained through pseudo single value decomposition \citep{yang2018design}. Reservoir computing with cellular automata (ReCA) \citep{yilmaz2014reservoir, nichele2017deep} is another family of models available in the library leveraging the package CellularAutomata.jl \citep{francesco_martinuzzi_2022_5879385}.


\paragraph{Training.} The training algorithms are implemented in a low level fashion, supporting all RC models of the library. Multiple training methods to obtain the output layer $\psi$ can be obtained from open source libraries such as MLJLinearModels.jl \citep{blaom2020flexible}, GaussianProcesses.jl \citep{gaussianprocesses.jl} and LIBSVM.jl \citep{libsvm}, a Julia porting of LIBSVM \citep{CC01a}. 

\paragraph{Prediction.} ReservoirComputing.jl leverages diverse prediction techniques. The generative approach allows the model to run autonomously: the predicted output $\mathbf{v}(t)$ is fed back into the model to obtain the prediction for the next time step $\mathbf{v}(t+1)$. The predictive approach instead uses the standard feature-label setup. 

\paragraph{States modifiers.} Other lower level implementations are included, such as the possibility to modify the state vectors. All the approaches detailed in \S \ref{theo} are included in the library.

\section{Code Quality}

The library is made available on Github, where it is continuously tested through Github actions. The package is part of the scientific machine learning (SciML) community \citep{sciml}, a NumFOCUS sponsored project since 2020 \citep{num}. An extensive online documentation is also provided with the library, where models are documented at length with multiple examples for different applications. 

\begin{figure}[t]
\centering
\includegraphics[scale=0.235]{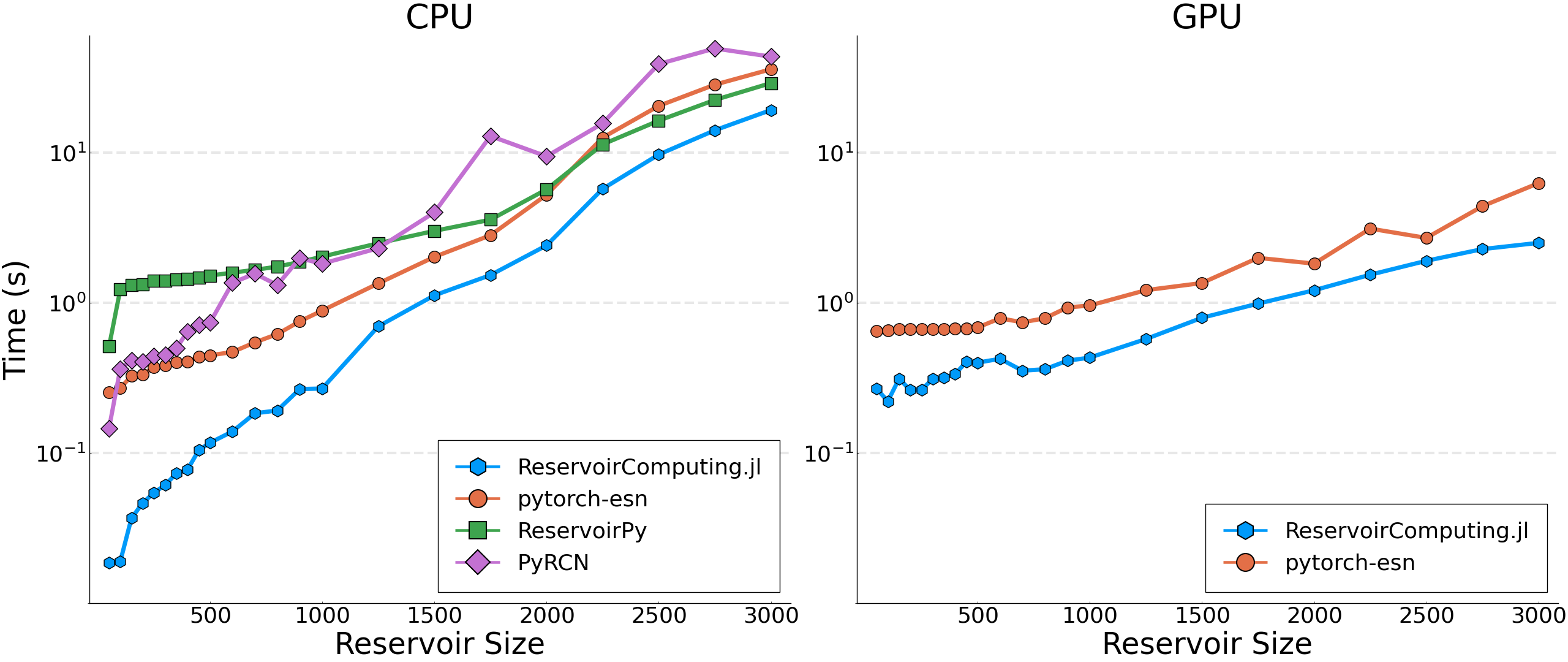}
\caption{Speed comparison of RC libraries.}
\label{fig: comp}
\end{figure}

A comparison of features and performance of ReservoirComputing.jl with similar libraries is provided in Table \ref{table: comp} and in Figure \ref{fig: comp}, respectively. The feature comparison in Table \ref{table: comp} showcases the large offering of models available. The plot in Figure \ref{fig: comp} illustrates the better computational speed of the library. The performance test task is a next step prediction of the Mackey-Glass system \citep{Glass2010} with time delay $\tau=17$. The dense reservoir matrix and the dense input matrix are generated with uniform distribution sampled from $[-1,1]$. The spectral radius of the reservoir matrix is scaled by 1.25. The ridge regression parameter is set to $10^{-8}$. Training and prediction lengths are both equal to 4999. The time reported is the sum of training and prediction. For central processing unit (CPU) computations the precision is set to float64, for the graphics processing unit (GPU) computations it is float32. EchoTorch is not present in the comparison as the version available on Github was not running on the examples provided. 
The versions tested are as follows: PyRCN v0.0.16, ReservoirPy v0.3.2, ReservoirComputing.jl v0.8 release candidate and pytorch-esn retrieved from Github on March 2022. All the simulations were run on a Dell XPS 9510 fitted with an Intel i9-11900H CPU, a Nvidia GeForce RTX 3050 Ti GPU and 16 GB of RAM.

\begin{table}[t]
\rowcolors{1}{lightgray}{} 
\begin{center}
\resizebox{\textwidth}{!}{%
\begin{tabular}{ l | c c c c c | c c c c c c | c c c c | c c | c c c |}
\toprule 
 \rowcolor{white} & \multicolumn{5}{c |}{Code quality} & \multicolumn{6}{c |}{Models: ESN} & \multicolumn{4}{c |}{Training} & \multicolumn{2}{c |}{Output} & \multicolumn{3}{c |}{Miscellanea} \\
 & \rotatebox[origin=c]{90}{Language} & \rotatebox[origin=c]{90}{Documentation} & \rotatebox[origin=c]{90}{API/Tutorials} & \rotatebox[origin=c]{90}{Tests} & \rotatebox[origin=c]{90}{GPU acceleration} & \rotatebox[origin=c]{90}{Leaky neurons} & \rotatebox[origin=c]{90}{Gated units} \rotatebox[origin=c]{90}{\citep{di2020gated}} & \rotatebox[origin=c]{90}{Hybrid ESN} \rotatebox[origin=c]{90}{\citep{pathak2018hybrid}} & \rotatebox[origin=c]{90}{Multiple reservoirs} & \rotatebox[origin=c]{90}{Deep architecture}\rotatebox[origin=c]{90}{\citep{gallicchio2018design}} & \rotatebox[origin=c]{90}{Non-ESN models} & \rotatebox[origin=c]{90}{Linear regression} & \rotatebox[origin=c]{90}{Gaussian regression} & \rotatebox[origin=c]{90}{SV regression} & \rotatebox[origin=c]{90}{Online training} & \rotatebox[origin=c]{90}{Generative} & \rotatebox[origin=c]{90}{Predictive} & \rotatebox[origin=c]{90}{States modifications} & \rotatebox[origin=c]{90}{Data sets} & \rotatebox[origin=c]{90}{Optimization}\\
 \midrule
 ReservoirComputing.jl & Julia & \color{green}\ding{51} & \color{green}\ding{51}/\color{green}\ding{51} & \color{green}\ding{51} & \color{green}\ding{51} & \color{green}\ding{51} & \color{green}\ding{51} & \color{green}\ding{51} & \color{green}\ding{51} & \color{green}\ding{51} & \color{green}\ding{51} & \color{green}\ding{51} & \color{green}\ding{51} & \color{green}\ding{51} & \color{red}\ding{55} & \color{green}\ding{51} & \color{green}\ding{51} & \color{green}\ding{51} & \color{red}\ding{55} & \color{red}\ding{55} \\
 ReservoirPy & Python & \color{green}\ding{51} & \color{green}\ding{51}/\color{green}\ding{51} & \color{green}\ding{51} & \color{red}\ding{55} & \color{green}\ding{51} & \color{red}\ding{55} & \color{red}\ding{55} & \color{red}\ding{55} & \color{green}\ding{51} & \color{green}\ding{51} & \color{green}\ding{51} & \color{red}\ding{55} & \color{red}\ding{55} & \color{green}\ding{51} & \color{green}\ding{51} & \color{green}\ding{51} & \color{red}\ding{55} & \color{green}\ding{51} & \color{green}\ding{51} \\
 EchoTorch & Python & \color{red}\ding{55} & \color{green}\ding{51}/\color{red}\ding{55} & \color{green}\ding{51} & \color{green}\ding{51} & \color{green}\ding{51} & \color{green}\ding{51} & \color{red}\ding{55} & \color{green}\ding{51} & \color{green}\ding{51} & \color{red}\ding{55} & \color{green}\ding{51} & \color{red}\ding{55} & \color{red}\ding{55} & \color{red}\ding{55} & \color{red}\ding{55} & \color{green}\ding{51} & \color{red}\ding{55} & \color{green}\ding{51} & \color{green}\ding{51} \\
 pytorch-esn & Python & \color{red}\ding{55} & \color{green}\ding{51}/\color{red}\ding{55} & \color{red}\ding{55} & \color{green}\ding{51} & \color{green}\ding{51} & \color{red}\ding{55} & \color{red}\ding{55} & \color{red}\ding{55} & \color{green}\ding{51} & \color{red}\ding{55} & \color{green}\ding{51} & \color{red}\ding{55} & \color{red}\ding{55} & \color{green}\ding{51} & \color{red}\ding{55} & \color{green}\ding{51} & \color{red}\ding{55} & \color{red}\ding{55} & \color{red}\ding{55} \\
 PyRCN & Python & \color{green}\ding{51} & \color{green}\ding{51}/\color{green}\ding{51} & \color{green}\ding{51} & \color{red}\ding{55} & \color{green}\ding{51} & \color{red}\ding{55} & \color{red}\ding{55} & \color{green}\ding{51} & \color{green}\ding{51} & \color{green}\ding{51} & \color{green}\ding{51} & \color{red}\ding{55} & \color{red}\ding{55} & \color{green}\ding{51} & \color{red}\ding{55} & \color{green}\ding{51} & \color{red}\ding{55} & \color{green}\ding{51} & \color{green}\ding{51} \\

\bottomrule
\end{tabular}}
\end{center}
\caption{Comparison of RC libraries. The \emph{code quality} section focuses on high level, quality of life library features. The subsections: \emph{Documentation} indicates whether the library includes general examples or how-to guides;  \emph{API/Tutorials} indicates whether API documentation and step by step tutorials are provided. In the ESN section we list common models included in the libraries. Subsection \emph{Multiple reservoirs} indicates whether native multiple reservoir matrix constructions are provided. In the training section, SV is short hand for support vector regression.  In the \emph{Miscellanea} section \emph{Data sets} indicates whether ready to use data sets are provided in the library; \emph{Optimization} stands for native hyperparameter optimization.}
\label{table: comp}
\end{table}

\section{Conclusion and Outlook}
ReservoirComputing.jl is a  comprehensive, modular and fast library for RC models with a strong focus on ESNs. In the future we plan to expand the model library by including LSMs and extreme learning machines (ELMs), while maintaining the intuitiveness of the current implementation.

\section*{Acknowledgements}
This work was partially supported by the German Federal Ministry of Education and Research (BMBF, 1IS18026B) by funding the competence center for Big Data and AI “ScaDS.AI” Dresden/Leipzig. K.M. is funded by the iDiv Flexpool program. M.D.M. and K.M. thank the European Space Agency for funding the "DeepExtremes" project (AI4Science ITT).

\bibliographystyle{unsrtnat}
\bibliography{rcjl}  






\end{document}